# An Integrated Approach Towards the Construction of an HCI Methodological Framework

**Tasos Spiliotopoulos**
Department of Mathematics & Engineering
University of Madeira
9000-390 Funchal, Portugal
tasos@m-iti.org

**Ian Oakley**
Madeira Interactive Technologies Institute
University of Madeira
9000-390 Funchal, Portugal
ian@uma.pt

**Abstract**
We present a methodological framework aiming at the support of HCI practitioners and researchers in selecting and applying the most appropriate combination of HCI methods for particular problems. We highlight the need for a clear and effective overview of methods and provide further discussion on possible extensions that can support recent trends and needs, such as the focus on specific application domains.

**Keywords**
User experience, usability, methods, framework, HCI, domains.

**ACM Classification Keywords**
H5.2. Information interfaces and presentation: User Interfaces - evaluation/methodology.

**General Terms**
Human factors.

**Introduction and background**
The use of human-centered design models and methodologies facilitates software development processes. However a plethora of approaches exists and it can be challenging for developers to appropriately match tools to problems. One way to address this issue is via methodological frameworks, which facilitate the development process by aiding selection of the most appropriate design methods according to the characteristics of a particular project.

The contribution of this paper is the presentation of a novel framework for systematically categorizing and evaluating HCI methods. It is anticipated that this framework will ease the process of selecting appropriate HCI methods for particular design and evaluation tasks. An overview of the framework is provided, its structure and qualities are discussed and directions for its future development are proposed. The discussion highlights areas of improvement for similar efforts, including providing effective overview of methods, and extensions that support the study of recent trends and needs in HCI, such as the shift of emphasis to user experience (UX) and the focus on specific application domains.

**Overview of methodological approaches**
An apparent way of categorizing HCI methods can be based on the development stage that the method is applied to, be it analysis, design, implementation, formative evaluation, or summative evaluation. However, such a categorization does not provide any direct appreciation of the different kinds of results and insights provided by a method. Nor does it highlight the resources required, or the fact that some methods can be effectively used in more than one development phase. Fitzpatrick and Dix proposed an alternative schema for categorizing HCI methods according to their strategic application [3]. In this approach, four strategies are proposed, based on the resources at human and system level (i.e. real or representative users, and real or representative system) thereby creating a 2 by 2 matrix with top level categories of real world, virtual engineering, soft modeling and hard review. However in this framework, methods are then classified at a second level based on their type and the way they are used, resulting ultimately in a usage analysis table. The result of this process is that the top-level categorization does not reflect the underlying goal of classifying the methods and, as such, does not offer useful insights. A third approach involves categorizing methods by the type of participants featuring in the UX evaluation [6]. This leads to a top-level breakdown into categories of lab tests, field studies, online surveys, and expert evaluations without actual users. Another recent approach classifies usability evaluation methods into data gathering & modeling methods, user interactions evaluation methods, and collaborative methods [4].

**An HCI methodological framework**
The first step in the development of the methodological framework presented in this paper was to gather a comprehensive corpus of HCI methods. Overall, 41 HCI methods were sourced from the literature and studied, analyzed and systematically described (for the complete list, see [8]). The set of methods included both traditional usability methods and those that take into account experiential aspects of a system.

In the framework described in this paper, a top-level categorization of the methods is achieved according to the way that they are used, resulting in four categories: *inquiry*, *prototyping*, *inspection* and *testing*, with 14, 8, 12 and 7 methods in each category, respectively.

Inquiry methods refer to the study of the use of an existing system or the characteristics of a system under construction by querying users and other stakeholders. Methods of this type typically require a considerable number of participants and, since they are based on participants' opinions and feelings about a system, provide subjective results. Examples include questionnaires, interviews and focus groups. Prototyping methods involve the construction of prototypes, which enable the study of the characteristics of a system at an early stage in the development process. Prototypes can be further classified according to their fidelity to the final system, level of detail, and scope, into high or low fidelity, and horizontal or vertical prototypes. Examples of prototyping methods include paper prototyping and storyboards. Inspection methods involve the close examination of a system by one or more HCI professionals. Typical inspections can be based on heuristics or guidelines, and can be driven by scenarios (i.e. walkthroughs). In testing methods, experiments on the system or a prototype are conducted with the participation of users. Typical examples include think-aloud protocol, wizard-of-Oz, and retrospective testing.

This categorization was selected as it highlights both the usage of a method and the type of resources that are required. In general, methods in different groups were found to exhibit high complementarity and low overlap in their results, focus and required resources. For example, inquiry methods tend to provide subjective results, as opposed to testing methods, whereas inspection methods do not require the direct participation of users, as inquiry and testing methods do.

The key to the methodological framework is the comparative evaluation of the HCI methods, which is based on a set of measurable criteria. This comparative evaluation, in combination with a good understanding of the workings of each method from their description and analysis, and a good apprehension of the needs of the system under development, can facilitate the selection of the most suitable combination of methods for each project.

The criteria for the comparative evaluation were selected so that they also enhance the general overview of the available methods. The HCI methods and evaluation parameters are tabulated in order to enable quick and effective overview and comparison. An excerpt of this table, comprising only 4 of the 41 methods studied is depicted in Table 1. The evaluation parameters are explained below.

- Type of method: This refers to the classification of methods as inquiry, prototyping, inspection or testing.
- Life-cycle stage: One or more of the following: requirements analysis, design, implementation, formative evaluation, and summative evaluation.
- Type of results obtained (quantitative - qualitative). This is a particularly important parameter, since methods that provide different types of results usually exhibit high complementarity and low overlap, thereby leading to a more efficient development process. Quantitative results are easily analyzed statistically, presented and interpreted in reports, and can be used for comparison of products or ideas in a straightforward way. On the other hand, qualitative results are not easily documented, but can provide important insights that can be easily missed in a quantitative approach.
- Bias (subjective - objective results). The results derived from the application of a method may be influenced to a significant extent by a personal predilection in part of a user or a usability expert.

**Table 1.** Excerpt from the HCI Methodological Framework depicting 4 of the 41 methods studied. The tabular presentation of the methods enables a quick and effective overview and comparison. The brief, descriptive analysis of the main strengths and weaknesses provides further insights aiding the selection of the most appropriate combination of methods.

| Name | Focus groups | Contextual interviews | Heuristic walkthrough | Automatic logging of use |
|---|---|---|---|---|
| Type | inquiry | inquiry | inspection | testing |
| Dev. phase | req. analysis, design, formative evaluation | req. analysis | design, formative and summative evaluation | formative and summative evaluation |
| Type of results | qualitative | qualitative | qualitative, quantitative | quantitative |
| Bias | subjective | subjective | objective | objective |
| Cost | low (recruitment of participants) | medium (recruitment of participants during work time, trip to the workplace) | low | low (recording and logging equipment, data analysis) |
| HCI experts | yes | yes | yes | no |
| No. of users | 6-9 | 5-10 | - | - (no users recruited specifically) |
| Level of detail | high, low | high, low | high, low | low |
| Immediacy | yes | yes | yes | no |
| Location | lab, work place | work place | lab | lab |
| Intrusiveness | yes | yes | no | no |
| Strengths | The users' preferences and ideas come from spontaneous reactions. Group dynamics come into effect. Easily repeatable. Can focus on specific issues. | Takes into account the context of use of a system. Can focus on specific issues and aspects of the system in detail. Most effective for exploring an application domain. | Inexpensive, flexible, structured, quick and repeatable evaluation method. Can be applied on low fidelity prototypes. Easily documented results. Can focus on specific parts of a system. | Demonstrates how a system is really used. Allows data collection from a large number of real users, in an automatic and systematic way. Allows for a longitudinal approach to studying users' behavior. Provides easily documented results. |
| Weaknesses | Results are subjective and not easily documented or analyzed. The presence of a group moderator is imperative to keep the group on track and make sure that participants do not influence each other. | Results are subjective and not easily documented or analyzed. Employee participants may be biased. | Heuristics limit significantly the scope of the evaluation. A walkthrough covers only a small part of the system. Inherent bias due to the selection of tasks/scenarios to be evaluated. | Provides answers to *how* people use a system, but not *why*. Does not take into account experiential aspects of the use of the system. The results require statistical analysis. Caution is needed as in not to breach users' privacy. |

This is something that should be taken into account in the interpretation of the results and the selection of methods.

- Cost. Includes items such as required equipment (e.g. a usability lab), prototype development costs, user recruitment costs and the cost of usability experts (if required).
- Need for recruiting HCI experts. Boolean parameter referring to whether the method requires HCI experts for correct execution.

- **Number of users.** A typical number of users that are needed to participate in the activities described in the method.
- **Level of detail of the results (high - low).** The results derived from the application of a method, may be low-level (e.g. relating to icon size or positioning) or high-level (e.g. general impressions of a system).
- **Immediacy of results.** Whether the method yields results immediately, or if further analysis (e.g. statistics) is required.
- **Location.** The site where a method's activities take place (e.g. lab, field, workplace).
- **Intrusiveness.** A user's behavior may be influenced by the presence of an observer, interviewer or recording equipment. This criterion highlights the extent to which the method is intrusive.
- **Strengths.** The strengths and main advantages of each method are described briefly using natural language.
- **Weaknesses.** The weaknesses and main disadvantages of each method are described briefly using natural language.

## Discussion

This framework integrates characteristics from similar efforts [3,6,7] to provide a complete and comprehensive catalogue and comparative tool.

The main advantages of the framework and the points of differentiation from similar attempts are enumerated below:

- **Systematic**. A lucid and eloquent overview of methods is achieved by categorizing and positioning them in a single table.
- **Critical Review**. Inclusion of a descriptive analysis of the main advantages and disadvantages (strengths and weaknesses) of each method. This brief overview will effectively assist HCI professionals in incorporating and applying a method in their projects.
- **Comprehensive**. The framework covers 41 HCI methods in total, embracing aspects of system development from initial conception to final testing. This compares well to previous attempts (e.g. [6,7]).
- **Extensible**. The framework is template-based, so it can be updated by appending new methods, and allows for the revision of the characteristics of each method in a collaborative way from the HCI community. New parameters and possible ways of categorization can be included to address emerging needs of the HCI community, such as the shift to UX or the focus on a specific domain.

The main disadvantages and limitations of the framework are as follows:

- **Subjective**. The characteristics and the values for each method's parameters have been elicited either from the relevant literature, or from the personal experience of a small number of HCI researchers and practitioners.
- **Non-experiential**. The framework does not explicitly address the experiential aspects of interaction with a system.

## Suggested enhancements

In order to address the shortcomings we propose a number of enhancements to the framework. These ideas can also prove to be useful enhancements to similar efforts undertaken in this area.

First, an interactive online version of the methodological matrix should be developed. This should make use of visualization elements and techniques to simplify the overview and review of methods and support the decision-making process. Drawing from examples of information visualization used in decision making (e.g. [1]), visual elements such as color, saturation, shape, size, texture and orientation can be used to convey information. Interactivity can support the decision making process by dynamically altering the visibility and visual elements in the online version of the framework. An online interactive version of the framework will further enhance the clear overview of the methods, which is a problem in large and detailed collections of methods (e.g. [4]), and is also expected to contribute to further disseminating the methods to the community of HCI practitioners, researchers, students, software developers and other stakeholders.

Second, an interactive online version of the matrix can be enhanced to support participation from HCI researchers and practitioners. This should yield similar benefits, in terms of exposure to a wide range of opinions, as those frameworks derived from survey data (e.g. [2,5,6,7]). However, integration into an online platform will be a more streamlined and efficient way, which is expected to contribute to the dissemination of the results. Of course, this is expected to be one step further than mere collaborative editing, as done in wikis, for example. The data from many participants, including some basic profile information, can be analyzed statistically and allow the dynamic provision of different views of the framework. For instance, it will be possible to find which methods can be used when one does not have access to users, or which methods are most applicable to a specific application domain.

**Acknowledgements**
The work reported in this paper is supported by FCT grant SFRH/BD/65908/2009.